\documentclass[%
 reprint,
 amsmath,amssymb,
 aps]{revtex4-1}

\usepackage{geometry,amsmath,amsthm}

\usepackage{graphicx}
\usepackage[colorlinks=true,linkcolor=black, citecolor=black,
urlcolor=black]{hyperref} 

\usepackage{mathrsfs}
\usepackage[usenames,dvipsnames]{xcolor}
\usepackage{color}
\usepackage{graphicx}%
\usepackage{dcolumn}%
\usepackage{bm}%
\usepackage{tikz}
\usetikzlibrary{calc}

\newcommand{\bal}{\begin{align}}
\newcommand{\eal}{\end{align}}
\newcommand{\beq}{\begin{equation}}
\newcommand{\eeq}{\end{equation}}
\newcommand\beqa{\begin{eqnarray}}
\newcommand\eeqa{\end{eqnarray}}
\newcommand\bea{\begin{array}}
\newcommand\eea{\end{array}}
\newcommand\comment[1]{{}}

\usepackage{youngtab}

    \newcommand{\COMMENT}[1]{}
    
    \newcommand{\neqa}{\nonumber\end{eqnarray}}

\def\[{\left[}
\def\]{\right]}

\def\[{\left[}
\def\]{\right]}

\def\<{\langle}
\def\>{\rangle}

\def\i2{\frac{i}{2}}

\def\be{\begin{eqnarray}}
\def\ee{\end{eqnarray}}
\def\no{\nonumber}

    \def\CM{{\cal M}}

    \def\<{\left\langle\,}
    \def\>{\, \right\rangle}
    \def\[{\left[}
    \def\]{\right]}

\newcommand{\wQ}{{\mathbb{Q}}}

   \def\GL{{\mathsf{GL}}}
   
   \def\SU{{\mathsf{SU}}}

    \def\N{\mathrm{N}}
    \def\M{\mathrm{M}}
   \def\es{\emptyset}

\begin{document}

\preprint{IPhT/t12/xxx}

 \title{Fast analytic solver of rational Bethe equations} 

\author{C.~Marboe$^{a,b}$, D.~Volin$^{a,c}$}
  
\affiliation{%
\(^a\)  School of Mathematics, Trinity College Dublin, College Green, Dublin 2, Ireland\\
\(^b\) Institut f\"{u}r Mathematik \& Institut f\"{u}r Physik, Humboldt-Universit\"{a}t zu Berlin,
Zum Gro\ss en Windkanal 6, 12489 Berlin, Germany\\
\(^c\) Bogolyubov Institute for Theoretical Physics, 14-b, Metrolohichna str. Kiev, 03680, Ukraine}

\begin{abstract}
In this note we propose an approach for a fast analytic determination of all possible sets of Bethe roots corresponding to eigenstates of rational $\GL(\N|\M)$ integrable spin chains of given not too large length, in terms of Baxter Q-functions. We observe that all exceptional solutions, if any, are automatically correctly accounted.

The key intuition behind the approach is that the equations on the Q-functions are determined solely by the Young diagram, and not by the choice of the rank of the $\GL$ symmetry. Hence we can choose arbitrary $\N$ and $\M$ that  accommodate the desired representation. Then we consider all distinguished Q-functions at once, not only those following a certain Kac-Dynkin path. 
\end{abstract}

\maketitle

\section{\label{sec:int}Introduction}
The spectral problem of integrable models is usually considered to be solved if one reduces it to the set of Bethe  equations \cite{Bethe:1931hc}. In the simplest case of the SU(2) XXX spin chain the equations read
\be\label{BAESU2}
\left( \frac{u_k+\frac \hbar 2}{u_k-\frac \hbar 2} \right)^L = - \prod_{j=1}^M \frac{u_k-u_j+\hbar}{u_k-u_j-\hbar}\,.
\ee
The Bethe equations for the $\GL(\N)$ case were formulated in \cite{Kulish:1983rd}, the $\GL(\N|\M)$  version was introduced in \cite{Kulish:1985bj}, and further studies were done in \cite{Tsuboi:1998ne,Ragoucy:2007kg} to include arbitrary choices of simple root systems and representations of spin chain nodes. 

One should however remember that the final numerical or analytic expressions for the energy are still not explicitly available, one should still solve the Bethe equations. After few initial attempts, one quickly realises that a direct brute force solution can be done in a very limited number of cases which is not satisfactory. Two typical scenarios where it is possible to proceed have been discussed in the literature. First, only a certain class of solutions may be of interest, e.g. the antiferromagnetic vacuum and some excitations around it. Then one avails tools to analyse these cases, especially in certain limits, like a thermodynamic limit, see e.g. \cite{Faddeev:1996iy}. Second, we could have a goal to generate numerical solutions. To this end one can deform the equations to the point where it is reasonably easy to solve them and then gradually remove the deformation. The deformation could be twists, inhomogeneities, changing length, strength of interaction etc., see e.g. \cite{Hao:2013jqa} for a recent successful effort in this direction. It is possible to reach truly spectacular results by numerics if we do not aim to immediately cover the whole spectrum but instead analyse different classes of solutions separately \cite{Bargheer:2008kj}. There is no doubt that the numerical machinery can be enhanced further upon need, however, probably, at the cost of a human effort in designing case-to-case algorithms.

In this note we propose a different approach for finding Bethe roots based on an extended set of functional equations. The method allows to generate  {\it all} possible physical solutions for not too large values of quantum numbers, and its particular strength is in the possibility to get explicit {\it analytic} answers which is unlikely to be possible in the above-discussed deformation-based techniques.

By an analytic answer we mean a Baxter polynomial 
\be\label{bQ}
\mathbb{Q}\equiv \prod_{j=1}^M (u-u_j)=u^M+\sum_{k=0}^{M-1}c^{(k)}u^k
\ee
with coefficients $c^{(k)}$ being explicit algebraic numbers, i.e. numbers of type $q_0+q_1\theta+q_2\theta^2+\ldots q_{K-1}\theta^{K-1}$, where $q_i$ are rational and $\theta$ is a root of a degree-$K$ polynomial equation with rational coefficients. $\theta$ may be expressible in radicals in the simplest cases.

Another advantage of the approach is that it is elementary from the point of view of human effort needed to use it. More precisely, it needs only well-developed and well-implemented functions in modern symbolic programming languages and hence the programming time is minimal.  Finally, it is not restricted to the $\SU(2)$ case, but it works, without modifications, for any $\GL(\N|\M)$ rational spin chain. 

The main disadvantage of our approach is that it directly targets finding all solutions at once and hence, naturally, it cannot produce results for too large values of the charges. On the other hand, when the charges are large, it is often irrelevant to find all solutions explicitly and hence other ways of treating Bethe equations should be preferred.

Section \ref{sec:rep} explains how to determine the number of solutions using representation theory. We present our solution algortihm in section \ref{sec:magic} and motivate it in section \ref{sec:behind}.

\section{Expected number of solutions} \label{sec:rep}

For the sake of simplicity, we will discuss only homogeneous chains in the fundamental (defining) representation of the $\GL$ group. Possible  generalisations are outlined in section~\ref{sec:Generalisations}. 

\begin{figure}
\setlength{\unitlength}{0.4mm}
\begin{picture}(120,85)

\color{gray}
\linethickness{.15mm}

\multiput(0,0)(0,15){3}{\line(1,0){110}}
\multiput(0,45)(0,15){2}{\line(1,0){30}}

\multiput(0,0)(15,0){3}{\line(0,1){65}}
\multiput(45,0)(15,0){5}{\line(0,1){30}}

\color{black}
\put(0,0){\vector(1,0){120}}
\put(0,0){\vector(0,1){75}}

\linethickness{0.5mm}

\put(0,0){\line(1,0){60}}
\put(0,15){\line(1,0){60}}
\put(0,30){\line(1,0){30}}
\put(0,45){\line(1,0){15}}

\put(0,0){\line(0,1){45}}
\put(15,0){\line(0,1){45}}
\put(30,0){\line(0,1){30}}
\put(45,0){\line(0,1){15}}
\put(60,0){\line(0,1){15}}

\put(-2,77){$a$}
\put(122,-2){$s$}

\end{picture}
\caption{\label{Fig:YoungThook}Example of $L=7$ and partition $(4,2,1)$ inscribed inside $2|2$ fat hook. $d_{\lambda}=\frac{7!}{6\cdot 4\cdot 2\cdot 3}=35$.}
\end{figure}
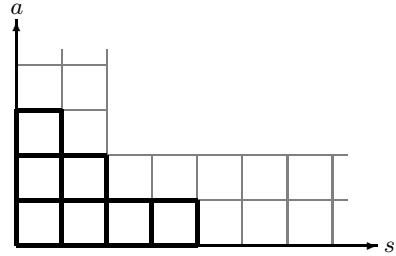

From the point of view of $\GL(\N|\M)$ action, the Hilbert space of a length-$L$ spin chain is the $L$'th tensor power ${\tiny\yng(1)}^{\otimes L}$. There is also  a natural action of the group $S_L$ permuting the nodes of the chain. The groups $\GL(\N|\M)$ and $S_L$ form a dual pair. Then, following the standard Schur-Weyl theory, one deduces the  decomposition of ${\tiny\yng(1)}^{\otimes L}$ into $\GL(\N|\M)$ irreps:
\be
{\tiny\yng(1)}^{\otimes L}=\bigoplus_\lambda d_{\lambda} \, V_{\lambda}\,,
\ee
where the sum runs over all possible integer partitions of $L$. We denote  isomorphism classes of $\GL(\N|\M)$ irreps  by $V_{\lambda}$. The multiplicity $d_{\lambda}$ of $V_{\lambda}$ occurrence is non-zero if the Young diagram depicting the partition $\lambda$ can be inscribed inside the $\N|\M$ fat hook, see Fig.~\ref{Fig:YoungThook}.  If it is non-zero then $d_{\lambda}$ is equal to the dimension of the dual $S_L$-representation, and it can be found by means of character theory or combinatorics. For instance, one can use the hook formula that reads
\be
d_{\lambda}=\frac{L!}{\prod\limits_{(a,s)\in \lambda}l_{a,s}}\,, \label{eq:hook}
\ee
where the product runs over all boxes of the Young diagram and $l_{a,s}$ is the length of the hook with the cusp at the box $(a,s)$.

A solution of the Bethe equations (or, more accurately, solutions of the below-defined Q-system) corresponds not to the one eigenstate of the spin chain, but to the whole multiplet $V_{\lambda}$. For the given partition
\be
\lambda\equiv (\lambda_1\geq \lambda_2\geq \lambda_3\geq\ldots)\,,\ \ \  \sum_{a}\lambda_a=L \quad
\ee
one should find precisely $d_{\lambda}$ solutions and this is our main goal.

The case $d_{\lambda}=0$ is of course not interesting. Absence of the corresponding representation is of purely "kinematical" origin stemming from the impossibility of organising the corresponding antisymmetrisations within the available flavour structure of the spin chain. 
On the other hand, the interesting case  of $d_{\lambda}\neq 0$ does not actually depend on $\N$ and $\M$ as long as kinematic restrictions are overcome. We will shortly see that the value of Q-functions does not depend on $\N$ and $\M$ either. Hence all $\GL(\N|\M)$  chains can be covered at once.

\section{\label{sec:magic}Magic recipe}
$\bullet$ Assign a monic polynomial in $u$ to each vertex $(a,s)$ of the Young diagram and denote it as $\wQ_{a,s}$. Its zeros shall be called Bethe roots, and the polynomial itself shall be called Baxter Q-function.

$\bullet$  Power counting for a spin chain in the fundamental representation: The degree of $\wQ_{a,s}$ is equal to the number of boxes of the Young diagram that are to the right and above the vertex. It is denoted as $M_{a,s}$. Explicitly, $M_{a,s}=L-\sum_{b=1}^{a} \lambda_b-\sum_{t=1}^{s}\lambda'_t+a\,s\,,$
where $\lambda'$ denotes the transposed diagram. 
Note that $M_{a,s}=0$ on the upper-right boundary of the diagram and hence $\wQ_{a,s}=1$ is known there.

$\bullet$  Boundary condition for homogeneous spin chain is $\wQ_{0,0}=u^L\,.$

$\bullet$
The polynomials should satisfy
\be
&&\wQ_{a+1,s}\wQ_{a,s+1}\propto \wQ_{a+1,s+1}^+\,\wQ_{a,s}^--\wQ_{a+1,s+1}^-\,\wQ_{a,s}^+\,.
\nonumber\\&&\nonumber\\
\label{mainQQ}
\ee
\vspace{-6mm}
\begin{figure}[h!]
\centering
\setlength{\unitlength}{0.25mm}
\begin{picture}(270,0)
\color{black}
\linethickness{0.5mm}
\put(0,5){\line(1,0){30}}
\put(0,35){\line(1,0){30}}
\put(0,5){\line(0,1){30}}
\put(30,5){\line(0,1){30}}

\put(40,5){\line(1,0){30}}
\put(40,35){\line(1,0){30}}
\put(40,5){\line(0,1){30}}
\put(70,5){\line(0,1){30}}

\put(90,5){\line(1,0){30}}
\put(90,35){\line(1,0){30}}
\put(90,5){\line(0,1){30}}
\put(120,5){\line(0,1){30}}

\put(130,5){\line(1,0){30}}
\put(130,35){\line(1,0){30}}
\put(130,5){\line(0,1){30}}
\put(160,5){\line(0,1){30}}

\put(180,5){\line(1,0){30}}
\put(180,35){\line(1,0){30}}
\put(180,5){\line(0,1){30}}
\put(210,5){\line(0,1){30}}

\put(220,5){\line(1,0){30}}
\put(220,35){\line(1,0){30}}
\put(220,5){\line(0,1){30}}
\put(250,5){\line(0,1){30}}

\normalsize
\put(0,35){\circle*{5}}
\put(70,5){\circle*{5}}
\put(120,35){\circle*{5}}
\put(130,5){\circle*{5}}
\put(210,35){\circle*{5}}
\put(220,5){\circle*{5}}

\put(111,28){\tiny$+$}
\put(201,28){\tiny$-$}
\put(132,9){\tiny$-$}
\put(222,9){\tiny $+$}

\put(75,18){$\propto$}
\put(165,18){$-$}
\put(33,18){$\cdot$}
\put(123,18){$\cdot$}
\put(213,18){$\cdot$}
\end{picture}
\end{figure}

We use $\propto$ to denote equality up to an overall normalisation.
We also use the notation $f^\pm(u)\equiv f(u\pm \hbar/2)$. The value of $\hbar$ can be made arbitrary by rescaling of $u$. Popular choices are $\hbar=\sqrt{-1}$ and $\hbar=1$.

Solutions of \eqref{mainQQ} by polynomials with provided constraints are in one-to-one correspondence with multiplets $V_{\lambda}$ that are eigenspaces of the spin chain. The  momentum and (what is typically considered as) energy are
\be
\label{energy momentum}
e^{ip}=\lim_{u\to 0}\frac{\wQ_{1,0}^+}{\wQ_{1,0}^-}\,,
\ \ \
E\propto \lim_{u\to 0}\partial_u\log \frac{\wQ_{1,0}^+}{\wQ_{1,0}^-}\,.
\ee
We wrote these expressions in a safe way, with $\lim_{u\to 0}$, to cover cases when $\wQ_{1,0}(\pm\hbar/2)=0$.

For a reader experienced with functional Bethe Ansatz   we readily note that our new message is not the well-known equation \eqref{mainQQ}, but to demand it for all Q-functions on the Young diagram {\it at once}. We  discuss the benefits of this point of view in the next section.\\

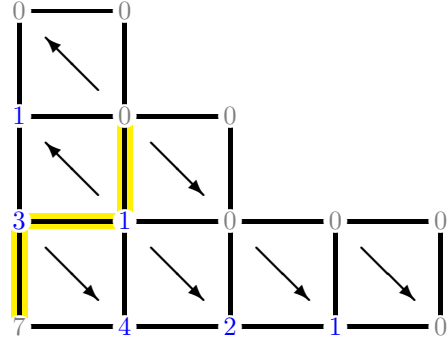
\begin{figure}[t]
\centering
\setlength{\unitlength}{0.35mm}
\begin{picture}(160,120)

\color{yellow}
\linethickness{2mm}
\put(0,0){\line(0,1){40}}
\put(0,40){\line(1,0){40}}
\put(40,40){\line(0,1){40}}

\color{black}
\linethickness{0.5mm}
\put(0,0){\line(1,0){160}}
\put(0,40){\line(1,0){160}}
\put(0,80){\line(1,0){80}}
\put(0,120){\line(1,0){40}}

\put(0,0){\line(0,1){120}}
\put(40,0){\line(0,1){120}}
\put(80,0){\line(0,1){80}}
\put(120,0){\line(0,1){40}}
\put(160,0){\line(0,1){40}}

\thicklines
\put(10,30){\vector(1,-1){20}}
\put(50,30){\vector(1,-1){20}}
\put(90,30){\vector(1,-1){20}}
\put(130,30){\vector(1,-1){20}}
\put(50,70){\vector(1,-1){20}}

\put(30,50){\vector(-1,1){20}}
\put(30,90){\vector(-1,1){20}}

\normalsize
\color{blue}
\put(0,120){\color{white}\circle*{9}}\put(-2.5,117){\color{gray}0}
\put(40,120){\color{white}\circle*{9}}\put(37.5,117){\color{gray}0}

\put(0,80){\color{white}\circle*{9}}\put(-2.5,77){1}
\put(40,80){\color{white}\circle*{9}}\put(37.5,77){\color{gray}0}
\put(80,80){\color{white}\circle*{9}}\put(77.5,77){\color{gray}0}

\put(0,40){\color{white}\circle*{9}}\put(-2.5,37){3}
\put(40,40){\color{white}\circle*{9}}\put(37.5,37){1}
\put(80,40){\color{white}\circle*{9}}\put(77.5,37){\color{gray}0}
\put(120,40){\color{white}\circle*{9}}\put(117.5,37){\color{gray}0}
\put(160,40){\color{white}\circle*{9}}\put(157.5,37){\color{gray}0}	

\put(0,0){\color{white}\circle*{9}}\put(-2.5,-3){\color{gray}7}
\put(40,0){\color{white}\circle*{9}}\put(37.5,-3){4}
\put(80,0){\color{white}\circle*{9}}\put(77.5,-3){2}
\put(120,0){\color{white}\circle*{9}}\put(117.5,-3){1}
\put(160,0){\color{white}\circle*{9}}\put(157.5,-3){\color{gray}0}	

\end{picture}
\caption{\label{fig:Young421} The values of $M_{a,s}$ in corresponding vertices. Yellow line is a chosen Kac-Dynkin path. Diagonal arrows point towards unknown Q-functions to be found recursively using \eqref{mainQQ}.}
\end{figure}

Now we introduce an algorithm generating explicit equations to be solved:

Choose a path on the diagram that connects the corner $(0,0)$ to a point on the upper-right boundary of the Young diagram. Typically it is more efficient to choose a path with the minimal number of Bethe roots along it. Write a generic polynomial ansatz for the Q-functions on the path: 
\be
\wQ_{a,s}=u^{M_{a,s}} +\sum_{k=0}^{M_{a,s}-1} c_{a,s}^{(k)}\, u^k\,.
\ee
Generate recursively the remaining Q-functions using \eqref{mainQQ}. Note that one will always be generating Q-functions as $\wQ_{\text{unknown}} \propto \frac{\wQ_a^+\wQ_b^--\wQ_a^-\wQ_b^+}{\wQ_c}$, i.e. one will not face the case when a finite-difference equation should be solved.

The unknown Q-function is a ratio of two polynomials, but it is required to be a polynomial itself, and thus it can be assigned to the quotient of the polynomial division. The remainder of this division should vanish. Hence we write:
\begin{align}
\label{quotient}
\wQ_{\text{unknown}} \propto {\rm Quotient}[\ {\wQ_a^+\wQ_b^--\wQ_a^-\wQ_b^+}\,,\,{\wQ_c}\ ]\,,&&
\\
\label{remainder}
0={\rm Remainder}[\ {\wQ_a^+\wQ_b^--\wQ_a^-\wQ_b^+}\,,\,{\wQ_c}\ ]\,.&&
\end{align}
Equation \eqref{quotient} allows one to promptly generate all the Q-functions explicitly in terms of $c_{a,s}^{(k)}$, one need not worry about validity of \eqref{remainder} for that. Thus we are able to work with all the Q-functions at once without introducing a zoo of auxiliary variables.

While generating all the Q-functions, we assemble in parallel a set of zero remainder conditions \eqref{remainder} which, in practice, become algebraic equations on $c_{a,s}^{(k)}$. Solving them is a question for standard pre-programmed routines in a symbolic programming language, we used {\it Mathematica}. Using the obtained equations proves to be significantly more efficient than using the Bethe equations like \eqref{BAESU2}. Also, the proposed procedure automatically properly accounts for the exceptional solutions which is a known issue with the Bethe equations.

As an example, we list the obtained timings for analytic solution of the $L=12$ $\SU(2)$ cases in {\it Mathematica} using a single core of a 3.2 GHz machine:
\begin{center}
\begin{tabular}{|c|c|c|c|c|c|c|c|}\hline
$M$ & 0& 1 & 2  & 3 & 4 & 5 & 6 \\ \hline
Solutions & 1 & 11 & 54 & 154 & 275 & 297 & 132  \\ \hline
Time (s) & 0.01 & 0.10 & 0.96 & 4.65 & 39 & 24& 18 \\ \hline
\end{tabular}
\end{center}
The treatment of the $M=5,6$ cases have small subtleties explained in appendix~\ref{sec:example}.

The {\it Mathematica} implementation of the algorithm outlined in this section is available in the ancillary files at arxiv.org. In this file, we analyse the efficiency of the method in more detail. The reader interested only in practical usage can skip the rest of the paper as it is devoted to explanation of the origins of the method.

\section{Behind the scenes} \label{sec:behind}
\subsection{Analysis of $\SU(2)$ case}

\paragraph{Shortcomings of Bethe equations.} When solving Bethe equations one should filter the obtained results. To commence, note that the solutions which differ only by permutation of Bethe roots correspond to the same physical state. Hence the only relevant information is concentrated in the coefficients $c^{(k)}$ of the Baxter polynomial, and we aim to find $c^{(k)}$ directly instead of the Bethe roots. 

There are also two potential singular scenarios when a more careful treatment is needed.

First, solutions with two or more coinciding Bethe roots have vanishing wave functions and should typically be discarded. However, by a fine tuning of parameters,  one can cook up a physical solution with coinciding Bethe roots (see e.g. Fig.~7 of \cite{Volin:2010xz}), so a physical case of coincident Bethe roots is at least conceivable and one should be a bit alert here to not miss a legit solution.

Second, there are so-called exceptional states when $\wQ^+$ and $\wQ^-$ have common zeros. Bethe equations suffer from division or multiplication by zero in this case and are hence ill-defined.

These issues can usually be resolved by introducing a regularisation, e.g. a twist, which can eventually be removed. However, this adds extra complexity to the practical analysis of the equations.

\paragraph{Baxter equation and  over-counting.}
Instead of Bethe equations, it is often better to use the Baxter equation \cite{Baxter:1972wg}
\be
(u^{+})^L\wQ^{--}+(u^{-})^L\wQ^{++}\propto T\,\wQ\,.
\ee


One demands that $T$ is  a polynomial in $u$. The  demand can be implemented, in style of section~\ref{sec:magic}, as conditions on the coefficients $c^{(k)}$:
\begin{align}\label{Remainder2}
0={\rm Remainder}[\ (u^{++})^L\wQ^-+(u^{--})^L\wQ^+,\wQ\ ]\,.
\end{align}
Equation \eqref{Remainder2} is equivalent to \eqref{BAESU2} in general situation but not in the above-mentioned singular cases. 

First, in the case of coinciding Bethe roots, it is stronger (the numerator should cancel the double zero of the denominator), so generically Q-functions with coinciding Bethe roots will not be allowed as solutions to the Baxter equation. To our knowledge, this properly resolves the issue of coinciding roots. 

Second, the remainder condition \eqref{Remainder2} does not suffer from singularities in the case of exceptional solutions. However, not all exceptional solutions of the Baxter equation are physical. Hence the Baxter equation has, generically, more solutions than is required by physics.
\paragraph{Q-system and right counting.}
The Baxter equation is a second-order finite-difference equation, hence it has another solution $\tilde\wQ$ which is linearly independent from $\wQ$ \cite{Pronko:1998xa}. The Wronskian determinant of the two solutions is
\be\label{bosQQ}
u^L\propto \wQ^+\tilde \wQ^--\wQ^-\tilde \wQ^+\,,
\ee
whereas $T$ can be found as $T\propto \wQ^{++}\tilde \wQ^{--}-\wQ^{--}\tilde \wQ^{++}$. The solution $\tilde\wQ$ is not guaranteed to be a polynomial but {\it it should be a polynomial} for all physical cases. This requirement automatically implies that $T$ is a polynomial and hence it implies \eqref{Remainder2}. However, it is stronger and it allows to discard wrong exceptional solutions. Solving \eqref{bosQQ} in polynomials generates precisely the physical solutions and only them \cite{Mukhin2009}.

Despite structural similarity, \eqref{bosQQ} is conceptually different from \eqref{mainQQ} as $u^L$ appears in two terms on the r.h.s. of \eqref{mainQQ} at $a=0,s=0$. In particular, we cannot use the remainder trick when solving \eqref{bosQQ} for $\wQ$ and $\tilde\wQ$. Supersymmetrisation is required to work with \eqref{mainQQ}.

\subsection{Supersymmetric Q-systems}
\paragraph{Q-system.}
Further developments in the spirit of Baxter's functional equation lead to the discovery of a plethora of QQ- and TQ-relations \cite{Krichever:1996qd}. The supersymmetric generalisation of these ideas was introduced in \cite{Tsuboi:1998ne} and coined to the form particularly relevant for us in \cite{Kazakov:2007fy}. 

We now sketch the B\"{a}cklund transform logic to introduce the notion of a Q-system. The reader is invited to consult \cite{Kazakov:2007fy} for a detailed discussion.

For a $\GL(\N|\M)$ system, one commences with a Q-function labelled as $Q_{12\ldots \N|12\ldots \M}$. The indices before $|$ are called bosonic indices and the indices after $|$ are called fermionic indices. One performs a special procedure (B\"{a}cklund transformation) that allows one to construct another Q-function, which has one index less, say $Q_{13\ldots \N|12\ldots \M}$. We repeat this procedure, in total $\N+\M$ times, until we are left with the Q-function which does not have indices at all: $Q_{\emptyset|\emptyset}$.

We can choose to remove indices in a different order, but the final result should not depend on the order chosen. For instance, removing first the index $\alpha$ and then the index $\beta$ should be equivalent to first removing $\beta$ and then $\alpha$. This puts constraints on the possible values of Q-functions -- the QQ-relations. 

If indices $\alpha$ and $\beta$ are of the same type (both bosonic or both fermionic) then the constraints are bosonic QQ-relations
\begin{align}\label{bosQQg}
Q_{A\alpha\beta|J}Q_{A|J}\propto Q_{A\alpha|J}^+Q_{A\beta|J}^--Q_{A\alpha|J}^-Q_{A\beta|J}^+\,.
\end{align}
Equation \eqref{bosQQ} is one example of \eqref{bosQQg} with $\wQ=Q_{\alpha|\es}$, $\tilde\wQ=Q_{\beta|\es}$ and $u^L=\wQ_{\es|\es}\wQ_{\alpha\beta|\es}$. 

If indices $\alpha$ and $\beta$ are of different type, the constraints are fermionic QQ-relations
\begin{align}\label{fermQQg}
Q_{A\alpha|J}Q_{A|J\beta}\propto Q_{A\alpha\beta|J}^+Q_{A|J}^--Q_{A\alpha\beta|J}^-Q_{A|J}^+\,.
\end{align}
Equation \eqref{mainQQ}  is an example of \eqref{fermQQg}. $\wQ_{a,s}$ is defined as the polynomial  $Q_{A|J}$ of the smallest degree among all Q-functions for which the number of indices in $A$ and $J$ are $a$ and $s$, respectively.  We can always adjust our choice of index labelling to have
\be\label{distQ}
\wQ_{a,s}=Q_{12\ldots a|12\ldots s}\,.
\ee
The full set of $2^{\N+\M}$ Q-functions supplemented with bosonic and fermionic QQ-relations shall be called Q-system.\\

The Q-system can also be viewed as a quantum spectral curve of the model in the following sense: Consider a classical curve defined by a characteristic equation $\det[\ \CM(u)-\lambda\ ]=0$ where $\CM$ is a $\N\times \N$ matrix (for simplicity we do only the bosonic case, a supersymmetric generalisation of the logic is possible). Quantise it by identifying $\lambda=e^{\hbar\partial_u}$ and getting a finite difference equation $\det[\ \CM-e^{\hbar\partial_u}\ ]Q=0$. It has $\N$ independent solutions -- Q-functions, in our notation it would be Q's with a single index. The Q-system is a collection of all possible Wronskian determinants built from these solutions.  Hence we see that a Q-system as an algebraic object is a quite generic set of relations, that can always be built once we have a rank-$N$ linear equation, so its emergence should not be particularly mysterious. Nevertheless, despite its simplicity, the system is very rich algebraically, especially the supersymmetric version, see \cite{Kazakov:2015efa}. 

What really  distinguishes the case of rational  integrable spin chains is that we know that {\it all} the Q-functions should be polynomials. Indeed, Q-functions were shown to be the eigenvalues of Baxter Q-operators  that were explicitly constructed for the rational spin chains \cite{Bazhanov:2008yc,Bazhanov:2010ts,*Bazhanov:2010jq,Kazakov:2010iu} and who are polynomials in  $u$.

\paragraph{Functional Bethe Ansatz and link to conventional Bethe equations.}
To determine the Q-system, it suffices to know only $\N+\M+1$ functions, all the rest can be obtained by QQ-relations. Two of these functions are already known: $Q_{\es|\es}=\wQ_{0,0}=u^L$ and $Q_{12\ldots \N|12\ldots \M}=\wQ_{\N,\M}=1$ set up the boundary data distinguishing a spin chain in the fundamental representation. An approach of functional Bethe Ansatz is to derive $\N-\M-1$ Bethe equations on the remaining $\N-\M-1$ unknowns, as we are going to review now.

 We already saw, in the $\SU(2)$ example, how a bosonic QQ-relation \eqref{bosQQ} yields the Baxter equation and then the Bethe equations. This immediately generalises to \eqref{bosQQg}, and the resulting {\it bosonic} Bethe equations read $\frac{Q_{A\alpha|J}^{++}Q_{A\alpha\beta|J}^-Q_{A|J}^-}{Q_{A\alpha|J}^{--}Q_{A\alpha\beta|J}^+Q_{A|J}^+}=-1$ at zeros of $Q_{A\alpha|J}$. Correspondingly, the {\it fermionic} Bethe equations following from \eqref{fermQQg} read $\frac{Q_{A|J}^{+}Q_{A\alpha\beta|J}^-}{Q_{A|J}^-Q_{A\alpha\beta|J}^+}=1$ at zeros of $Q_{A\alpha|J}$.

Now one chooses one particular chain of B\"{a}cklund transformations and write $\N+\M-1$ Bethe equations containing only the Q-functions of this chain. It is often convenient (though not compulsory) to constrain the possible choices  and work with chains comprised solely from  the distinguished Q-functions $\wQ_{a,s}$ defined by \eqref{distQ}. These particular B\"{a}cklund transformations can be suggestively depicted by a two-dimensional  path \cite{Kazakov:2007fy}; on this path, we move one step down every time a bosonic index is removed and one step left every time a fermionic index is removed. It is the same path as in Fig.~\eqref{fig:Young421}, with the only difference that the choice of the beginning point is anchored to the position ($\N,\M$).

The path is naturally bijected to the Kac-Dynkin diagram (turns of the path are fermionic nodes) and the Bethe equations for the zeros of Q-functions on the path are the Bethe equations defined by the Cartan matrix of the corresponding simple root system, in spirit of \cite{Reshetikhin:1987bz}. We refer to \cite{Kazakov:2007fy} for a more detailed discussion.

Working with Bethe equations has above-discussed shortcomings, hence it is more accurate to work directly with those QQ-relations that imply the Bethe equations. Even so, using QQ-relations only along the Kac-Dynkin path {\it does not} always guarantee that all the $2^{N+M}$ Q-functions are polynomials. In general position, the polynomiality can be  advocated by techniques known as duality transformations \cite{Woynarovich,*tJmodel,Tsuboi:1998ne,Pronko:1998xa,Gromov:2007ky}, however the logic behind these techniques has loopholes in the case of exceptional solutions which indeed allow singularities to emerge. For instance, for the path in Fig.~\ref{fig:Young421}, Bethe equations follow from two fermionic QQ-relations. These two relations do not guarantee regularity of the whole Q-system, which can be demonstrated by explicit counter-examples.

Hence, in general, using Bethe Ansatz logic, even in its form of QQ-relations will lead to overcounting. Therefore, once we get a solution of Bethe equations, we still need to generate all other Q-functions and check whether they are polynomials.

\subsection{Supersymmetric extension trick}
Because we need to generate other Q-functions anyway to check their polynomiality, let us come back to \eqref{bosQQg} and \eqref{fermQQg} and impose all of these conditions from the very beginning. It  is particularly easy to implement fermionic QQ-relations because of the quotient/remainder trick while there is no such advantage for the bosonic QQ-relations.

The good news is that it is possible to completely abandon bosonic relations, even for originally non-supersymmetric spin chains! To this end we consider the states of our departing $\GL(\N|\M)$ spin chain as special states inside a chain with larger $\GL(\tilde \N|\tilde \M)$ symmetry. This allows us to use Q-functions $Q_{A|J}$ with $A$ and $J$ being strings of integers of arbitrarily long desired length. In particular we can work with $\wQ_{a,s}$ on the Young diagram. Now we use the following lemma proved in appendix~\ref{sec:lemma}:\\

{\bf Polynomiality lemma.} {\it Let $\wQ_{a,s}$ be polynomials for any $(a,s)$ on the Young diagram and $\wQ_{a,s}=1$ on its boundary. Then all Q-functions of the $\GL(\N|\M)$ Q-system are polynomials if $(\N,\M)$ is a point on the boundary of the diagram.}

The practical outcome is: We need to ensure the polynomiality of $\wQ_{a,s}$ only, and these distinguished Q-functions are related solely by fermionic QQ-relations \eqref{mainQQ}. This is the essence of the recipe of section~\ref{sec:magic}. 

Example: in the $\SU(2)$ case, we consider the states of this chain as a particular class of states of the $\GL(2|L-M)$ chain and identify $\wQ\equiv\wQ_{1,0}=Q_{1|\emptyset}$ and $\tilde\wQ\equiv Q_{2|\emptyset}$. Note that $\wQ_{2,0}=Q_{12|\emptyset}=1\neq Q_{2|\emptyset}$. We never work with $\tilde\wQ$ in the proposed formalism, but we know that it is a polynomial as long as $\wQ_{a,s}$ are polynomials. Further details about this example are given in appendix~\ref{sec:example}.\\

Let us comment on why restriction to the Young diagram emerges. The degree of Q-functions  is dictated by representation theory. The upper boundary of the Young diagram is the place where the degree $M_{a,s}$ of $\wQ_{a,s}$ drops to zero. If the boundary of the Young diagram does not touch the inner boundary of $\N|\M$ hook then this is an atypical representation of $\GL(\N|\M)$ subject to shortening conditions. The $\GL(\N|\M)$ Q-system is affected by the shortenings, some of its Q-functions are zero \cite{Kazakov:2015efa}; however one can identify a $\GL(\N'|\M')$ subsystem of non-zero Q-functions  where $(\N',\M')$ is on the boundary of the Young diagram.


\subsection{Completeness and minimality questions}

\paragraph{Completeness.}
Any physical state is represented by a polynomial Q-system. Hence we are certain that any physical state would appear among solutions of the proposed set of equations. But can we still overcount, i.e. are there any non-physical polynomial solutions? At this stage we don't have a complete proof,  with the exception of the $\SU(2)$ case \cite{Mukhin2009}. However it is easy to check the status for every explicit choice of the integer partition $\lambda$. In all cases where solutions could be generated, the number of solutions precisely equaled $d_{\lambda}$. We hence have a solid evidence that no overcounting occurs and conjecture that solutions of \eqref{mainQQ} are in bijection with irreducible multiplets of spin chains, with {\it no need} of regularising the cases of exceptional solutions.

There are a number of interesting ways to approach a full proof of completeness: One can $q$-deform the system and then go to the crystal basis limit where one can count using rigged configurations \cite{KKR}. Another option is to introduce a twist and to make an analysis like in section 3.3 of \cite{Kazakov:2015efa}. Finally, we can work towards generalisation of the results in \cite{Mukhin2009}. All these options have several interesting adjacent combinatorial aspects. We leave their exploration for a future work. 

\paragraph{Minimal set of equations.}
The reader might have spotted that the function $\wQ_{0,3}$ in the example of Fig.~\ref{fig:Young421} is always polynomial provided the other $\wQ_{a,s}$ are polynomials. Hence the requirement of polynomiality of all $\wQ_{a,s}$ on the Young diagram is definitely not the minimal possible one. The interesting question is: What is the minimal requirement that ensures polynomiality of the Q-system?

We conjecture that one needs to require only polynomiality of $\wQ_{a,s}$ with $0\leq a,s\leq \N_{\rm min}$, where $\N_{\rm min}|\N_{\rm min}$ is the smallest possible symmetric (with $\N=\M$) fat hook that accommodates the Young diagram. We found severe constraints that support this hypothesis and could even prove it in some simple cases; also we checked it on explicit examples and found counter-examples showing that choosing a smaller minimal set is, generically, not enough. However we are not able to present a clean proof due to conceptually possible very subtle arrangements of zeros of Q-functions in exceptional solutions.

Note that chasing minimality is not necessarily beneficial.  Adding extra equations may improve computation efficiency even if they are devoid of new information.  In real computations, we actually tinker with choosing between the conjectured minimal, or even sub-minimal, set and the maximal set of equations \eqref{mainQQ} to optimise performance. Sometimes extreme choices are the best and sometimes intermediate ones are. For instance, generation of the $M=6$, $L=12$ states of the $\SU(2)$ spin chain is optimised and correctly works for the choice of $\GL(2|4)$ supersymmetric extension, which is in the mid-way between the minimal  $\GL(2|2)$ extension and the $\GL(2|6)$ extension required by the polynomiality lemma. For this particular case, we did prove that using only $\GL(2|2)$ extension is theoretically enough, however  {\it Mathematica} simply failed to solve the minimal set of equations.

\section{\label{sec:Generalisations}Generalisations}
The presented approach can be straightforwardly generalised to describe arbitrary inhomogeneities and to have spin chain sites in arbitrary representation of the symmetry algebra. This is achieved by modifications of the polynomial structure imposed on the Q-system. E.g. inhomogeneities $\chi_i$ are introduced through the replacement $\wQ_{0,0}=u^L\to \prod_{i=1}^L(u-\chi_i)$. The most general polynomial ansatz to account for all options is explicitly known \cite{Kazakov:2007fy}. Appendix C of \cite{Kazakov:2015efa} provides this ansatz in notations directly used in the current paper, except for the fact that a Hodge-dual picture is presented, with $Q_{\es|\es}=1$. To use the dual picture, one should define $\wQ_{a,s}=Q_{\overline{12\ldots a}|\overline{12\ldots s}}$ instead of \eqref{distQ}. The bar here means the complementary set.

Furthermore, the approach works also for spin chains with non-compact symmetries of $\SU(\N,\N'|\M)$ type. This generalisation will be published separately, as a part of our ongoing project \cite{spec} of explicit computation of the planar AdS/CFT spectrum using the quantum spectral curve \cite{Gromov:2013pga,*Gromov:2014caa}.

 \section{Conclusions and discussion} 
 In this paper we proposed an approach to solving Bethe equations which has a new shifted emphasis from the commonly used Lie algebra and Kac-Dynkin diagram perspective to the QQ-relations themselves naturally accommodated by a Young diagram. This approach turned out to be an efficient practical tool, but it also provides us with several lessons and suggests new directions of research.
 
 First of all, the emergence of Young diagrams clearly highlights the combinatorial nature of the diagonalisation procedure. It is worth to recall here that the Hamiltonian and other conserved charges are elements of the group algebra $\mathbb{C}[S_L]$. Apparently, the eigenvalue problem is more closely related to the permutation group rather than the $\GL(\N|\M)$ symmetry. Indeed, we have demonstrated that the final outcome is insensible to $\N$ and $\M$. Further exploration of these observations should merge this research with the work of \cite{Mukhin:2010uw}.
 
 Then, the fact that the Bethe equations can be solved explicitly up to lengths $L\simeq 12$, despite the obtained algebraic numbers being very complex and bulky expressions, is still quite a mystery for us. This clearly indicates the need of exploring the Galois group structure of the obtained equations to search for further hidden symmetries. One can add here a quite plausible fact: provided the completeness conjecture is true, the fermionic QQ-relations \eqref{mainQQ} on the Young diagram is a way to perceive the spectrum of a physical system as points of an algebraic variety and hence the methods of algebraic geometry are welcomed now. 

Another lesson is that supersymmetrisation is an important practical tool, to study even non-supersymmetric systems. Quite similar tricks can be applied also to simplify the derivation of the thermodynamic Bethe Ansatz equations \cite{Volin:2010xz} and to classify unitary representations of non-compact groups \footnote{M.Gunaydin, D.Volin, to appear}. Each of the QQ-relations \eqref{mainQQ} central for our construction is actually a functional Bethe Ansatz for the $\GL(1|1)$ spin chain. We get a clear hint that the structure of integrable systems/symmetry algebras can be understood to a large extent or even completely by supersymmetrising them and studying the behaviour of small building blocks that have $\GL(1|1)$ symmetry. In physical terms, we should seek to reduce the problem to the one of free fermions in sufficiently large auxiliary space.

 \begin{acknowledgments}
D.V. would like to thank IHES and IPhT, C.E.A.-Saclay for hospitality, where part of this work was done.  The work of C.M. was partially supported by the People Programme (Marie Curie Actions) of the European Union's Seventh Framework Programme FP7/2007-2013/ under REA Grant Agreement No 317089 (GATIS). 
\end{acknowledgments}
 
 \appendix
 
  \section{\label{sec:example}$\SU(2)$ example}
 
Our goal is to find all $M$-magnon solutions of length $L=12$ $\SU(2)$ spin chain. There should be ${L \choose M} - {L \choose M-1}$ of them. The corresponding Young diagram is given by the partition $(\lambda_1,\lambda_2)=(L-M,M)$.

On the boundary of the Young diagram, one has $\wQ_{2,s}=1$ for $0\leq M$ and $\wQ_{1,s}=1$ for $M\leq s\leq L-M$.

The polynomial $\wQ_{1,0}$ is of degree $M$. It is nothing but $\wQ$ defined in \eqref{bQ}. Its zeros are the Bethe roots solving \eqref{BAESU2}. Our goal is to determine the coefficients $c^{(k)}$. 

Polynomials $\wQ_{1,s}$ for $0\leq s\leq M$ are recursively found by 
\begin{align}
\wQ_{1|s+1} \propto \frac{\wQ_{2,s+1}^+\wQ_{1,s}^--\wQ_{2,s+1}^-\wQ_{1,s}^+}{\wQ_{2,s}}\propto \wQ_{1,s}^+-\wQ_{1,s}^-\,.
\end{align}
There are no zero remainder conditions generated here.

The nontrivial remainder requirement comes from the following recursion: One sets $\wQ_{0,0}=u^L$ and then
\be\label{eq:4}
\wQ_{0|s+1} \propto \frac{\wQ_{1,s+1}^+\wQ_{0,s}^--\wQ_{1,s+1}^-\wQ_{0,s}^+}{\wQ_{1,s}}\,,
\ee
and use the quotient/remainder trick explained in the main text. Note that the degree of $\wQ_{1,s}$ is $M-s$. Hence $\wQ_{1,s}$ with higher $s$ will generate simpler remainder conditions. This is one of the reasons that make the discussed approach efficient. The results were summarised at the end of section~\ref{sec:magic}.

This computation has two subtleties. One is that the brute-force use of {\it Mathematica} in $M=6$ case yields 152 solutions. A closer analysis shows that this happens because {\it Mathematica} fails to recognise that some of its generated solutions are identical. To improve the situation, we considered a smaller extension $\GL(2|4)$ instead of $\GL(2|6)$ and hence used only those relations \eqref{mainQQ} that involve $\wQ_{a,s}$ with $s\leq 4$. This proved to be beneficial, and we recovered the correct number of 132 solutions.

To solve the $M=5$ case, we had to supplement the system of equations with the momentum quantisation condition which reads, on the level of Q-functions, $\wQ(\hbar/2)^L-\wQ(-\hbar/2)^L=0$. Note that it follows from \eqref{mainQQ}, but adding it explicitly as an extra equation helps.

Using the momentum quantisation works as follows: Factorise the polynomial $\wQ(\hbar/2)^L-\wQ(-\hbar/2)^L$ over the field of rational numbers,
\be
&&x^6-y^6=
\\ \nonumber
&&(x-y)(x+y)(x^2-x y+y^2)(x^2+ x y +y^2)\,,
\ee
where $x=\wQ(\hbar/2)$ and $y=\wQ(-\hbar/2)$. Then, pick each of the four factors one by one and, equating it to zero, use this as an extra constraint to add to the system of equations. After getting solutions for all four cases, polish the result by removing duplicates (which are those with $\wQ(\hbar/2)=\wQ(-\hbar/2)=0$ thus appearing in each of the factor choices).

 \section{\label{sec:lemma}Proof of polynomiality lemma}

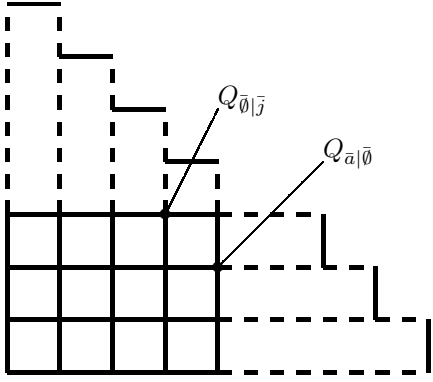
\begin{figure}[t]
\centering
\setlength{\unitlength}{0.35mm}
\begin{picture}(160,140)
\color{black}
\linethickness{0.1mm}
\qbezier(80,40)(80,40)(120,80)
\qbezier(60,60)(60,60)(80,100)

\color{black}
\linethickness{0.5mm}
\put(0,0){\line(1,0){80}}
\put(0,20){\line(1,0){80}}
\put(0,40){\line(1,0){80}}
\put(0,60){\line(1,0){80}}

\put(0,0){\line(0,1){60}}
\put(20,0){\line(0,1){60}}
\put(40,0){\line(0,1){60}}
\put(60,0){\line(0,1){60}}
\put(80,0){\line(0,1){60}}

\put(0,140){\line(1,0){20}}
\put(20,120){\line(1,0){20}}
\put(40,100){\line(1,0){20}}
\put(60,80){\line(1,0){20}}

\multiput(0,60)(0,10){8}{\line(0,1){5}}
\multiput(20,60)(0,10){8}{\line(0,1){5}}
\multiput(40,60)(0,10){6}{\line(0,1){5}}
\multiput(60,60)(0,10){4}{\line(0,1){5}}
\multiput(80,60)(0,10){2}{\line(0,1){5}}

\put(160,0){\line(0,1){20}}
\put(140,20){\line(0,1){20}}
\put(120,40){\line(0,1){20}}

\multiput(80,0)(10,0){8}{\line(1,0){5}}
\multiput(80,20)(10,0){8}{\line(1,0){5}}
\multiput(80,40)(10,0){6}{\line(1,0){5}}
\multiput(80,60)(10,0){4}{\line(1,0){5}}

\normalsize
\put(120,82){$Q_{\bar{a}|\bar{\emptyset}}$}
\put(80,102){$Q_{\bar{\emptyset}|\bar{j}}$}

\color{black}
\put(80,40){\circle*{4}}
\put(60,60){\circle*{4}}

\end{picture}
\caption{Separation of Young diagram into an $N\times M$ rectangle, upper wing and right wing.}
\label{fig:NNsq}
\end{figure}

Consider an arbitrary Young diagram and assume that all distinguished $Q$-functions are polynomial. The diagram can be decomposed, in several ways, into an $N\times M$ rectangle that touches the boundary of the diagram and two wings above and to the right of the rectangle, see Fig.~\ref{fig:NNsq}.

On the $N\times M$ rectangle, the $Q$-function in the upper right corner is $Q_{1...N|1...M}\equiv Q_{\bar{\emptyset}|\bar{\emptyset}}\equiv \wQ_{N,M} = 1$. Let us assume that  $Q_{\bar{a}|\bar{\emptyset}}$ and $Q_{\bar{\emptyset}|\bar{j}}$ are polynomials for all $a,j$. Then $Q_{\bar{a}|\bar{j}}$ is a polynomial as well. Indeed, it satisfies
\be
Q_{\bar{a}|\bar{\emptyset}}Q_{\bar{\emptyset}|\bar{j}}\propto Q_{\bar{a}|\bar{j}}^+-Q_{\bar{a}|\bar{j}}^-\,,
\ee
an equation which clearly has a polynomial solution. This solution is denoted as $Q_{\bar{a}|\bar{j}}\propto\Psi^-(Q_{\bar{a}|\bar{\emptyset}}Q_{\bar{\emptyset}|\bar{j}})$, see \cite{Marboe:2014gma,Kazakov:2015efa} for a detailed discussion of $\Psi$-operator properties.

 In this proof, an overall shift of the spectral parameter $u$ is of no importance, hence in the following we will abundantly replace $\Psi^{\pm}$ with $\Psi$ in order to avoid bulky expressions.

All other $Q$-functions on the rectangle are special determinant combinations of $Q_{\bar{a}|\bar{\emptyset}}$, $Q_{\bar{\emptyset}|\bar{j}}$, and $Q_{\bar{a}|\bar{j}}$ \cite{Kazakov:2015efa,Tsuboi:2011iz}. This means that if $Q_{\bar{a}|\bar{\emptyset}}$ and $Q_{\bar{\emptyset}|\bar{j}}$ are polynomial, all other $Q$-functions on the rectangle will be.

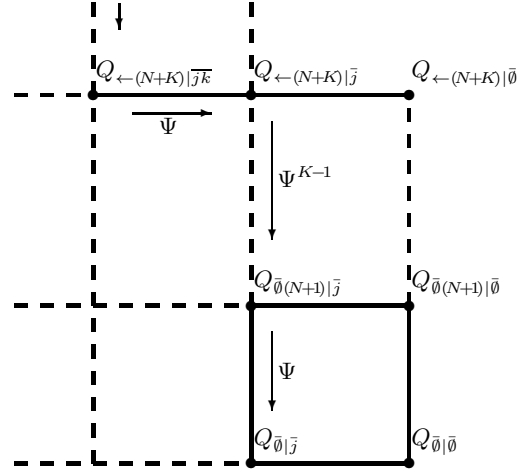
\begin{figure}[t]
\centering
\setlength{\unitlength}{0.35mm}
\begin{picture}(180,175)

\linethickness{0.5mm}

\put(90,0){\line(1,0){60}}
\put(90,60){\line(1,0){60}}
\put(90,0){\line(0,1){60}}
\put(150,0){\line(0,1){60}}

\multiput(0,0)(10,0){9}{\line(1,0){5}}
\multiput(0,60)(10,0){9}{\line(1,0){5}}
\multiput(0,140)(10,0){3}{\line(1,0){5}}

\multiput(30,0)(0,10){18}{\line(0,1){5}}
\multiput(90,60)(0,10){12}{\line(0,1){5}}
\multiput(150,60)(0,10){8}{\line(0,1){5}}

\put(30,140){\line(1,0){120}}

\put(91,7){$Q_{\bar{\emptyset}|\bar{j}}$}
\put(151,7){$Q_{\bar{\emptyset}|\bar{\emptyset}}$}
\put(151,67){$Q_{\bar{\emptyset}(\!N\!+\!1\!)|\bar{\emptyset}}$}
\put(91,67){$Q_{\bar{\emptyset}(\!N\!+\!1\!)|\bar{j}}$}

\put(90,0){\circle*{4}}
\put(150,0){\circle*{4}}
\put(90,60){\circle*{4}}
\put(150,60){\circle*{4}}

\put(151,147){$Q_{\leftarrow(\!N\!+\!K\!)|\bar{\emptyset}}$}
\put(91,147){$Q_{\leftarrow(\!N\!+\!K\!)|\bar{j}}$}
\put(31,147){$Q_{\leftarrow(\!N\!+\!K\!)|\overline{jk}}$}

\put(30,140){\circle*{4}}
\put(90,140){\circle*{4}}
\put(150,140){\circle*{4}}

\thinlines
\put(98,50){\vector(0,-1){30}}
\put(100,32){$\Psi$}

\put(98,130){\vector(0,-1){45}}
\put(100,105){$\Psi^{K\!-\!1}$}

\put(45,133){\vector(1,0){30}}
\put(55,125){$\Psi$}

\put(40,175){\vector(0,-1){10}}

\end{picture}
\caption{Relations between $Q$-functions on upper wing.}
\label{fig:uw}
\end{figure}

As the two wings behave identically, it is sufficient to make the argument on one wing. We here focus on the upper wing to show that $Q_{\bar{\emptyset}|\bar{j}}$ are polynomial. Denote the height of the outermost right column in the wing by $K$.  Noting that $Q_{\bar{\emptyset}|\bar{\emptyset}} \propto Q_{\bar{\emptyset}(\!N\!+\!1\!)|\bar{\emptyset}}\propto1$, we can generate $Q_{\bar{\emptyset}|\bar{j}}$ purely in terms of $Q_{\bar{\emptyset}(\!N\!+\!1\!)|\bar{j}}$ through \eqref{mainQQ}:
\be
Q_{\bar{\emptyset}(\!N\!+\!1\!)|\bar{j}}&\,\propto\,&Q_{\bar{\emptyset}|\bar{j}}^- - Q_{\bar{\emptyset}|\bar{j}}^+ \no\\ 
\Leftrightarrow \quad Q_{\bar{\emptyset}|\bar{j}}&\, \propto\,& \Psi \!\left(Q_{\bar{\emptyset}(\!N\!+\!1\!)|\bar{j}}\right)\,,
\ee
thus preserving polynomiality. We can continue in this way until we reach the end of the column:
\be
Q_{\bar{\emptyset}|\bar{j}} &\,\propto\,& \Psi^K\!\left(Q_{\leftarrow(\!N\!+\!K\!)|\bar{j}}\right)\,,
\ee
where the notation $(\leftarrow n)\equiv (12...n)$ is used.

Now, $Q_{\leftarrow(\!N\!+\!K\!)|\bar{M}}\propto 1$ as it is on the inner boundary. 
Together with $Q_{\leftarrow(\!N\!+\!K\!)|\bar{j}}$, $j\!\neq\! M$, it forms a bosonic $QQ$-relation \eqref{bosQQg} with $Q_{\leftarrow(\!N\!+\!K\!)|\bar{\emptyset}}\propto 1$ and $Q_{\leftarrow(\!N\!+\!K\!)|\overline{jM}}$:

\be
Q_{\leftarrow(\!N\!+\!K\!)|\overline{jM}}&\, \propto \,&Q_{\leftarrow(\!N\!+\!K\!)|\bar{j}}^- - Q_{\leftarrow(\!N\!+\!K\!)|\bar{j}}^+ \no\\ \quad  \Leftrightarrow \quad Q_{\bar{\emptyset}|\bar{j}} &\,\propto\,& \Psi^{K+1}\!\left(Q_{\leftarrow(\!N\!+\!K\!)|\overline{jM}}\right).\quad\quad
\ee

For $j\!=\!M\!-\!1$, the argument is a distinguished $Q$-function, and we have thus proven that $Q_{\bar{\emptyset}|\overline{(\!M\!-\!1\!)}}$ is polynomial. For $j\!\le\! M\!-\!2$, we have rewritten $Q_{\bar{\emptyset}|\bar{j}}$ in terms of a multiple of polynomialty preserving $\Psi$-operations on $Q_{\leftarrow(\!N\!+\!K\!)|\overline{jM}}$. See Fig. \ref{fig:uw} for an overview. We can now consider what is left of the second column from the right (denote the height by $K_2$) in exactly the same way as we treated the first. By exactly the same argument as above, we  find that $Q_{\bar{\emptyset}|\overline{(\!M\!-\!2\!)}}$ is polynomial, while the rest are related by $\Psi$-operations to $Q_{\leftarrow(\!N\!+\!K_2\!)|\overline{j(\!M\!-\!1\!)M}}$. By recursion, we then see that all $Q_{\bar{\emptyset}|\bar{j}}$ are related to distinguished $Q$-functions by multiple $\Psi$-operations, so if all distinguished $Q$-functions are polynomial, then it follows that all Q-functions on the $N\times M$ rectangle are polynomial as well. 

Finally recall that the $N\times M$ rectangle can be chosen arbitrarily as long as it touches the boundary of the Young diagram $\qed.$ \\

\vfill

%
%


\bibliography{bibliography}

\end{document}